# Photometric and kinematical analysis of Koposov 12 and Koposov 43 open clusters

**W. H. Elsanhoury[1, 2]**

[1]Physics Department, Faculty of Science and Arts, Northern Border University, Rafha Branch, Saudi Arabia. Email: elsanhoury@nbu.edu.sa and welsanhoury@gmail.com

[2]Astronomy Department, National Research Institute of Astronomy and Geophysics (NRIAG) 11421, Helwan, Cairo, Egypt, Affiliation ID: 60030681

ORCID iD: https://orcid.org/0000-0002-2298-4026

**Abstract**

We present a photometric and kinematical analysis of two and poorly studied open clusters; Koposov 12 (FSR 802) and Koposov 43 (FSR 848) by using cross-matched data from PPMXL and Gaia DR2 catalog. We use astrometric parameters to identify 285 and 310 cluster members for Koposov 12 and Koposov 43, respectively. Using the extracted member candidates and isochrone fitting to near-infrared (J, H, $K_s$) and Gaia DR2 bands (G, $G_{BP}$, $G_{RP}$), and Color Magnitude Diagrams (CMDs), we have estimated ages: log (age/yr) = 9.00 ± 0.20 and 9.50 ± 0.20, and distances d = 1850 ± 43 pc and 2500 ± 50 pc for Koposov 12 and Koposov 43, respectively, assuming Solar metallicity (Z=0.019). The estimated masses of the cluster derived using initial mass function and synthetic CMD are 364 ± 19 $M_\odot$ and 352 ± 19 $M_\odot$. We have also computed their velocity ellipsoid parameters based on (3 × 3) matrix elements $(\mu_{ij})$.

**Keywords:** Open clusters: Koposov 12 (FSR 802) and Koposov 43 (FSR 848) – PPMXL and Gaia DR2 – Photometry: color-magnitude diagrams – Stars: luminosity and mass functions – Kinematics: dynamical evolution, velocity ellipsoid parameters.

## 1. Introduction

Open star clusters are interesting targets as they give vital information on stellar structure, kinematics, and evolution of the Galactic disc (Yadav *et al.* 2011). The present paper is a part of our continuing series (Elsanhoury and Nouh 2019) whose goal is to obtain the main astrophysical and kinematical properties of open clusters considering crossmatch between near-infrared region (NIR) J, H, $K_s$ due to positions and proper motions on the International Celestial Reference Frame (ICRS) system (PPMXL catalog, Roeser *et al.* 2010) and Gaia photometric system G, $G_{BP}$, $G_{RP}$ (Gaia DR2 catalog, Gaia Collaboration 2018).

Since a new interstellar extinction in the near-infrared region (NIR) J, H, and $K_s$ give an opportunity to profoundly burrow in the spiral arm regions where most of the open clusters are concentrated. This availability of huge near-infrared region (NIR) surveys allows researchers to perform an automated search to investigate new clusters (Yadav *et al.* 2011). Utilizing accessible information due to (PPMXL) by combining United States Naval Observatory (USNO-B1.0, Monet *et al.* 2003) all-sky catalog completeness down to V = 21 and the 2 μm all-sky survey catalog (2MASS, Skrutskie *et al.* 2006) got to be confirmed by coordinating utilizing later catalog like Global Astrometric Interferometer for Astrophysics Data Release 2 (Evans *et al.* 2018). Gaia data gives recent reliable distances and kinematics to a large number of cluster members with greater accuracy. These recently available space parameters are very crucial and give clues to cluster disruption and the build-up of the field population (Fürnkranz *et al.* 2019). Recent publications making use of second Gaia data have provided membership lists for over a thousand clusters, however, many nearby objects listed in the literature have so far evaded detection (Cantat-Gaudin and Anders 2020).

The catalog of the second Gaia data release comes to a G-band magnitude of 21 (9 magnitudes fainter than Tycho-Gaia Astrometric Solution (TGAS, Michalik *et al.* 2015). At its faint end, the Gaia DR2 astrometric precision is accurate with that of Tycho-Gaia Astrometric Solution (TGAS), whereas for stars brighter than (G ≲ 15) the precision is about ten times way better than in Tycho-Gaia Astrometric Solution (TGAS), thus allowed to extend membership determinations to fainter and more distant objects (like our clusters under investigations). For more than 1.3 billion sources, Gaia DR2 catalog presents astrometric five parameter solutions; central coordinates, proper motion in right ascension and declination and parallaxes ($\alpha, \delta, \mu_\alpha \cos\delta, \mu_\delta, \pi$), moreover the magnitudes in the three passbands of the Gaia photometric system G, $G_{BP}$, $G_{RP}$ with precisions at the mmag level. Thus leads to analyze dynamical and kinematical evolutions.

Koposov *et al.* (2008) looking for Galactic star clusters in large multiband surveys to find new star clusters. Glushkova *et al.* (2010) recorded 168 newly open clusters, among which 26 are embedded ones. From this list we have chosen two clusters; Koposov 12 (FSR 802) with diameter of 9 arcmin; $(\alpha, \delta)_{2000} = (06^h\ 00^m\ 56^s.20,\ 35^d\ 16^m\ 36^s.00)$ and (l, b) = $(176^o.17014,\ 06^o.01963)$ and Koposov 43 (FSR 848) with diameter of 8 arcmin; $(\alpha, \delta)_{2000} = (05^h\ 52^m\ 14^s.60,\ 29^d\ 55^m\ 09^s.00)$ and (l, b) = $(179^o.92431,$ 01°.73987) (Koposov *et al.* 2008), in arrange to subject them to a photometric and kinematical investigation.

In this paper, we aim to understand the photometric and kinematical structure of these two open clusters. We have re-estimated their cluster parameters (reddening, distance modulus, ages) concerned with J, H, $K_s$, G, $G_{BP}$, and $G_{RP}$ photometry using membership selected based on full astrometric solution (utilizing proper motions and the magnitude uncertainties). On the other hand, our study of these two open clusters is the first one to estimate their kinematical and dynamical properties, into which we have computed their velocity ellipsoid parameters (VEPs) based on spatial velocities estimation (U, V, W) and the matrix elements $(\mu_{ij})$, the projected distances $(X_\odot, Y_\odot, Z_\odot)$, and the Solar elements $(S_\odot, l_A, b_A, \alpha_A, \delta_A)$.

This paper is organized in such a way that; Section 2 describes the data used in this study for estimation centers and surface density distribution. Section 3 deals with age, reddening, and distance. The next Section shows the luminosity and mass functions. Section 5 deals with the relaxation and internal motion processes of these open clusters due to the study of their dynamical and kinematical structure. Finally, the conclusion of this work is displayed in Section 6.

## 2. Data analysis

In the present work, we have used the fundamental parameters of these two open clusters inferred by Sampedro *et al.* (2017), Koposov *et al.* (2008), and the Milky Way Star Clusters project (MWSC, Kharchenko *et al.* 2013); which is based on (2MASS, Skrutskie *et al.* 2006) photometry and (PPMXL, Roeser *et al.* 2010) astrometry. The parameters are listed in Table 1.

In our purpose of analysis, we have crossmatched two sources of data. One concerned with the second intermediate Gaia Data Release (Gaia DR2) for row data collected within the first 22 months of nominal mission processed by the Gaia Data Processing and Analysis Consortium (DPAC). Gaia DR2 is setting a new major achievement for Gaia's mission in stellar, Galactic, and extra Galactic studies. It provides position, trigonometric parallax, radial velocity, and proper motions in both directions for more than 1 billion stars, as well as three broad bands photometric magnitudes; G (330 – 1050 nm), the Blue Prism $G_{BP}$ (330 – 680 nm), and Red Prism $G_{RP}$ (630 – 1050 nm) for sources brighter than 21 mag (Weiler 2018).

**Table 1:** The fundamental parameters of the two open clusters Koposov 12 (FSR 802) and Koposov 43 (FSR 848).

| Parameters | Koposov 12 (FSR 802) | Koposov 43 (FSR 848) | References |
|---|---|---|---|
| $\alpha$ | $06^h\ 00^m\ 55^s.99$ | $05^h\ 52^m\ 15^s.00$ | Sampedro *et al.* (2017) |
| | $06^h\ 01^m\ 02^s.64$ | $05^h\ 52^m\ 17^s.40$ | Kharchenko *et al.* (2013) |
| $\delta$ | $35^d\ 16^m\ 36^s.01$ | $29^d\ 55^m\ 09^s.01$ | Sampedro *et al.* (2017) |
| | $35^d\ 16^m\ 37^s.20$ | $29^d\ 53^m\ 49^s.20$ | Kharchenko *et al.* (2013) |
| $l$ | $176^o.159$ | $179^o.901$ | Sampedro *et al.* (2017) |
| | $176^o.170$ | $179^o.924$ | Kharchenko *et al.* (2013) |
| $b$ | $05^o.999$ | $01^o.744$ | Sampedro *et al.* (2017) |
| | $06^o.020$ | $01^o.740$ | Kharchenko *et al.* (2013) |
| Distance (pc) | 2000 | 2800 | Sampedro *et al.* (2017) |
| | 1900 | 3000 | Kharchenko *et al.* (2013) |
| E(B-V) (mag) | 0.510 | 0.380 | Sampedro *et al.* (2017) |
| | 0.450 | 0.650 | Kharchenko *et al.* (2013) |
| log (age/yr) | 8.78 | 9.30 | Sampedro *et al.* (2017) |
| | 8.190 | 9.115 | Kharchenko *et al.* (2013) |
| (m-M) (mag) | 11.55 ± 0.03 | 12.21 ± 0.09 | Koposov *et al.* (2008) |
| Diameter (arcmin) | 9.00 | 8.00 | Koposov *et al.* (2008) |
| $R_v$ (km/s) | 13.6 ± 1.90 | 26.6 ± 0.90 | Kharchenko *et al.* (2013) |
| $\mu_\alpha$ (mas/yr) | 0.699 ± 0.009 | -0.037 ± 0.021 | Cantat-Gaudin *et al.* (2018) |
| $\mu_\delta$ (mas/yr) | -1.732 ± 0.011 | -1.664 ± 0.014 | Cantat-Gaudin *et al.* (2018) |

The second source is devoted here with the PPMXL catalog (Roeser *et al.* 2010); which determines the mean positions and proper motions on the (ICRS) system by combining USNO-B1.0 and 2MASS astrometry. PPMXL (Roeser *et al.* 2010) contains about 900 million objects, some 410 million with 2MASS photometry (Roeser *et al.* 2010). 2MASS had drawn photometric passband observations of the sky for millions of galaxies and nearly a half-billion stars (Carpenter 2001) simultaneously NIR regions; J (1.25 $\mu$m), H (1.65 $\mu$m), and $K_s$ (2.17 $\mu$m) with sensitivity; J (5.8 mag), H (15.1 mag), and $K_s$ (14.3 mag) bands at S/N=10.

### 2.1 Cluster center determination

To start our calculations and to download complete row data, we utilized Gaia DR2[1] source. Koposov 12 (FSR 802) and Koposov 43 (FSR 848) open clusters are located near the Galactic plane; |b| < 6º; Koposov 12 (FSR 802) and |b| < 2º; Koposov 43 (FSR 848), have diameters less than 10 arcmin. To identify the extent of the cluster from the background stellar density, we use radial density profile (RDP), where we performed star count over an extracted data within 10 arcmin. The data includes angular distances (in arcmin) from the center, right ascension (in degrees), and declination (in degrees) with epoch 2015.0 about 3147 stars; Koposov 12 (FSR 802) and 4164 stars; Koposov 43(FSR 848) open clusters are downloaded.

---

[1] https://vizier.u-strasbg.fr/viz-bin/VizieR?-source=I/345

We performed binning along the right ascension ($\alpha$) and declination ($\delta$), with a bin size of 1.00 arcmin (Maciejewski and Niedzielski, 2007; Maciejewski *et al.* 2009) by two opposite strips were cut along ($\alpha, \delta$). Figure 1 shows the resulting Histogram which was built along those ($\alpha, \delta$) and the two Gaussian curves fittings are applied to the profiles of star counts in right ascension ($\alpha$) and declination ($\delta$) respectively. Table 2 presents our estimated values of the new centers (i.e. maximum peaks) for both clusters respectively, including the mean (average) ($\mu$) with the standard errors is taken to be $\pm 1\sigma$ (i.e. standard deviation) of the Gaussian distribution function f(x), i.e. $f(x) = (1/\sqrt{2\pi})\exp[-(x-\mu)^2/2\sigma^2]$. With such results, we may conclude that:

- Our new results of right ascension ($\alpha$) for Koposov 12 (FSR 802) bounded between Sampedro *et al.* (2017) and Kharchenko *et al.* (2013), where our estimation is greater by about $01^s.61$ that given by Sampedro *et al.* (2017) and smaller by about $05^s.04$ with Kharchenko *et al.* (2013). Also, our new estimation of declination ($\delta$) is greater by about $19^s.19$ and $18^s.00$ for those both authors, respectively.
- On the other hand, the comparison may be accomplished with other Koposov 43 (FSR 848) cluster, into which our new estimation of right ascension ($\alpha$) is smaller by about $00^s.12$ and $02^s.52$ with Sampedro *et al.* (2017) and Kharchenko *et al.* (2013), respectively. Other than that, our declination ($\delta$) is smaller by about $04^s.21$ that given with Sampedro *et al.* (2017), and is greater by about $01^m 15^s.60$ with Kharchenko *et al.* (2013).

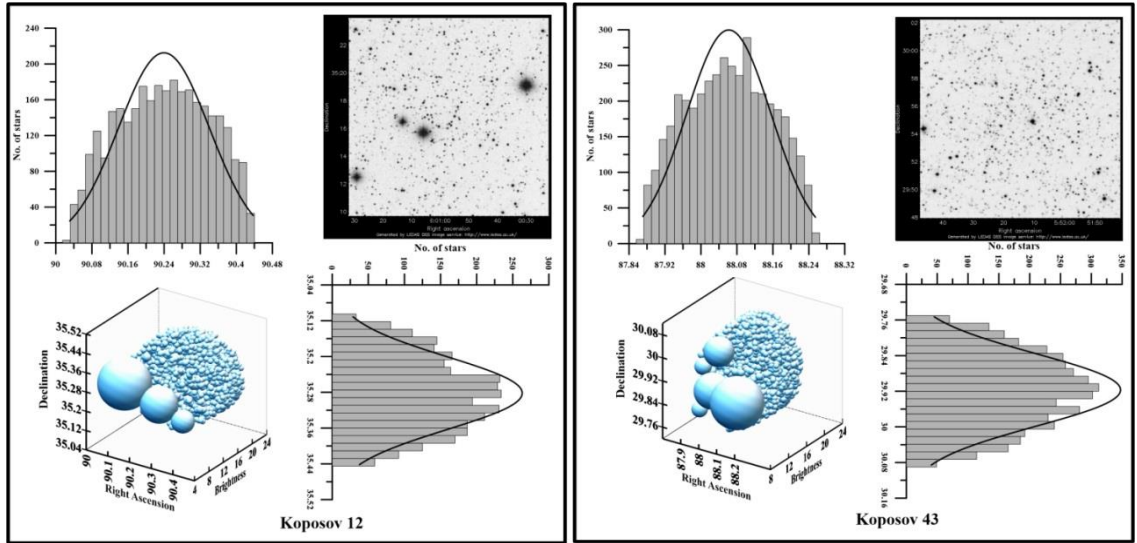

**Fig. 1:** The images are taken from LEicester Database and Archive Service (LEDAS) Digitized Sky Survey (DSS[2]) of Koposov 12 (FSR 802) and Koposov 43 (FSR 848) in the *left; (upper right)* and *right; (upper right) panels* respectively. On the other hand, showing the 3D density of stars on the sky with Gaia DR2 of both clusters; *left; (lower left)* and *right; (lower left) panels* respectively. The Gaussian fits (solid lines) provide the coordinates of highest density areas in ($\alpha$) and ($\delta$) for both clusters; *diagonally of both panels*, respectively.

**Table 2:** Our new center estimation for Koposov 12 (FSR 802) and Koposov 43 (FSR 848) open clusters.

| Parameters | Koposov 12 (FSR 802) (3147 stars) | Koposov 43 (FSR 848) (4146 stars) |
|---|---|---|
| Max. peak (Ra.) $_{degrees}$ | 90.240 ± 0.101 | 88.062 ± 0.095 |
| Max. peak (Dec.) $_{degrees}$ | 35.282 ± 0.082 | 29.918 ± 0.082 |
| $\alpha$ | $06^h\ 00^m\ 57^s.60$ | $05^h\ 52^m\ 14^s.88$ |
| $\delta$ | $35^d\ 16^m\ 55^s.20$ | $29^d\ 55^m\ 04^s.80$ |
| $l^o$ | $176^o.148$ | $179^o.922$ |
| $b^o$ | $6^o.007$ | $1^o.741$ |

## 2.2 The radial density profile

We downloaded a new worksheet data: 3173 stars; Koposov 12 (FSR 802) and 4162 stars; Koposov 43 (FSR 848), with our new center estimation appeared above in Table 2, we apply the King (1962) profile with annular bins; 0.90 arcmin (Koposov 12) and 1.00 arcmin (Koposov 43), to estimate their structural parameters; i.e. core radius ($r_{core}$), central surface density ($f_o$), and background surface density ($f_{bg}$);

$$\rho(r) = f_{bg} + \frac{f_o}{1 + (r/r_{core})^2}, \tag{1}$$

Figure 2 shows the King model fit and its uncertainties (estimated using OriginPro package) for the density distribution of these two open clusters. In expansion, we will characterize the limiting radius ($r_{lim}$) (Tadross and Bendary, 2014) as the radius at which the line represents the value of the background density intersects the King model fitting curve. At this point the background star density $(\rho_b = f_{bg} + 3\sigma_{bg})$, where $(\sigma_{bg})$ is fluctuation (uncertainty) of background surface density ($f_{bg}$). Mathematically, we have:

$$r_{lim} = r_{core} \sqrt{\frac{f_o}{3\sigma_{bg}} - 1}. \tag{2}$$

---

[2] https://www.ledas.ac.uk/DSSimage

**Table 3:** Our results of the internal structure for Koposov 12 (FSR 802) and Koposov 43 (FSR 848) with other published one.

| Parameters | Koposov 12 (FSR 802) | Koposov 43 (FSR 848) | References |
|---|---|---|---|
| $f_{bg}$ (stars arcmin$^{-2}$) | 0.069 ± 0.016 | 0.093 ± 0.006 | Present study |
| $f_o$ (stars arcmin$^{-2}$) | 0.921 ± 0.111 | 0.951 ± 0.028 | Present study |
| $r_{core}$ (pc) | 0.805 ± 0.002 | 0.571 ± 0.003 | Present study |
|  | 0.663 | 0.681 | Kharchenko *et al.* (2013) |
| $r_{lim}$ (pc) | 3.466 ± 0.240 | 3.946 ± 0.213 | Present study |
|  | 3.48 | 5.24 | Kharchenko *et al.* (2013) |
| $\delta_c$ | 14.324 ± 3.785 | 11.226 ± 3.351 | Present study |
| C | 4.305 ± 0.482 | 6.911 ± 0.380 | Present study |

In what takes after we provided a microscopic investigation of our internal cluster structure estimation, counting $r_{core}$, $f_{bg}$, and $f_o$, where numerical values of these parameters are listed in Table 3. We may take note that, our calculated core radius ($r_{core}$) is higher than by approximately 0.142 pc and smaller with 0.110 pc for both clusters respectively as compared with Kharchenko *et al.* (2013). Our limiting radius ($r_{lim}$) is smaller than by almost 0.014 pc and 1.294 pc for those clusters respectively, as compared with Kharchenko *et al.* (2013) database.

On the other hand, for our case study; it first one to compute the density contrast parameter $(\delta_c)$ $(i.e. \delta_c = 1 + f_o/f_{bg})$ like in Table 3. Interestingly, the density contrast parameter $(\delta_c)$ comes to high values $(7 \leq \delta_c \leq 23)$ (Bonatto and Bica, 2009) as expected from compact star clusters.

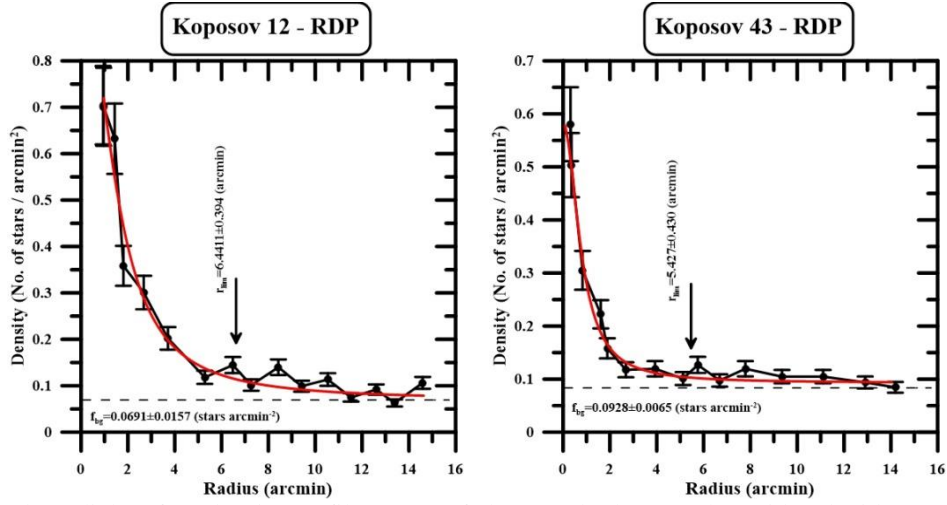

**Fig. 2:** The radial surface density profile (RDP) of clusters; the *left panel* considered with Koposov 12 (FSR 802), and the *right panel* presents that for Koposov 43 (FSR 848) with error bars. The fitted solid lines were applied with King's model, and the dashed lines represent the background field density $f_{bg}$.

The last row of Table 3, gives the concentration parameter (C) of these both clusters, which concerned the ratio between limiting and core radii (i.e. $C = r_{lim} / r_{core}$). Nilakshi *et al.* (2002) concluded that the angular size of the coronal region is almost ($6r_{core}$). Whereas Maciejewski and Niedzielski (2007) detailed, that ($r_{lim}$) extended between ($2r_{core} - 7r_{core}$). Our concentration parameter (C) is in a good agreement with Maciejewski and Niedzielski (2007) which about $4.305 \pm 0.482$ and $6.911 \pm 0.380$ for both clusters respectively.

Now, our ultimate objective is to determine the space density (stars arcmin$^{-3}$). Consider for the stellar system (e.g. star clusters) with radius (R), the luminosity per unit volume at position $r(r < R)$, shows that the surface brightness I(R) and the luminosity density j(r) are related by the formulae (Binney and Termaine, 2008) like;

$$I(R) = 2 \int_R^\infty dr \frac{rj(r)}{\sqrt{r^2 - R^2}}, \qquad (3)$$

Where

$$j(r) = -\frac{1}{\pi} \int_R^\infty \frac{dR}{\sqrt{R^2 - r^2}} \frac{dI}{dR}. \qquad (4)$$

The classical space density distribution (surface brightness per unit volume) in a specific direction of the Galactic plane of a given kind of association (e.g. star clusters) is usually determined with Blaauw and Schmidt (1965). Here the volume element (i.e. density) is determined by dividing the cluster into shells with certain width depending on the number of stars in each radius interval (shell), then we divided the counts in each shell by the volume (zone) of this shell (see van Rhijn 1936; Elsanhoury *et al.* 2011; Elsanhoury 2020).

To compute the star density (stars arcmin$^{-3}$) and according to Freedman and Diaconis's (1981) rule, we can divide these above remarks of n = 3173 stars; Koposov 12 (FSR 802) and 4162 stars; Koposov 43 (FSR 848) into 16 and 17 groups, respectively, according to their radial distances (arcmin). Statistical steps in constructing a frequency distribution of these points under consideration are shown in Table 4, into which:

- The number (k) of intervals can be determined with, k = (max. distance – min. distance / h), where (h) is the bin width.
- The bin width h with Freedman and Diaconis (1981) is h = 2 IQR/n$^{1/3}$, where (n) is the sample size, and (IQR) is the sample interquartile range.

**Table 4:** The statistical parameters of Koposov 12 and Koposov 43 open clusters according to Freedman and Diaconis's (1981) rule.

| Parameters | Koposov 12 (FSR 802) | Koposov 43 (FSR 848) |
| --- | --- | --- |
| n | 3173 | 4162 |
| Min. distance (arcmin) | 0.1818 | 0.1365 |
| Max. distance (arcmin) | 9.9997 | 9.99959 |
| IQR | 4.707 | 4.834 |
| h | 0.641 | 0.601 |
| k | 16.00 | 17.00 |

Table 5 presents the intervals (classes) into which first and second columns give the intervals (arcmin) and the center of the interval (arcmin) respectively, the third column presents the frequencies (counts), and the last column gives the density (stars arcmin$^{-3}$). Figure 3 shows our development volume density (stars arcmin$^{-3}$) profile distribution of both clusters, respectively.

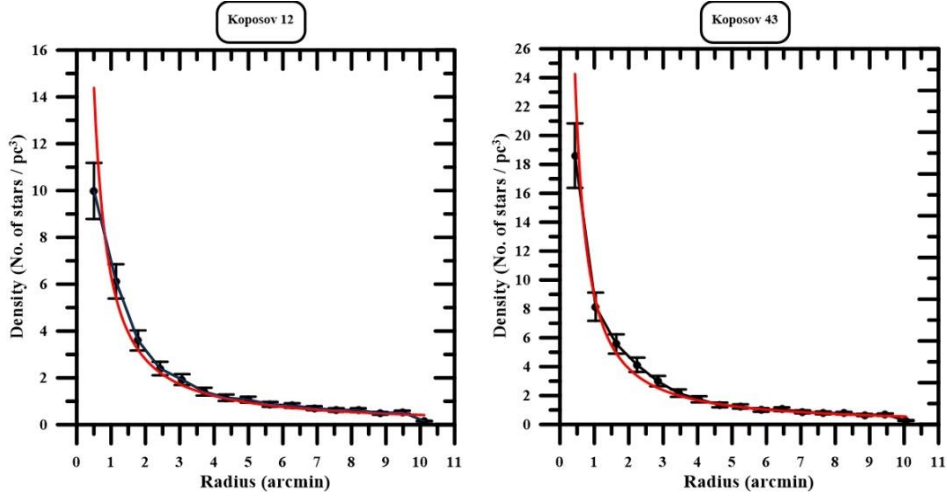

**Fig. 3:** The volume density profile (stars arcmin$^{-3}$) of Koposov 12 (FSR 802); *left panel* and Koposov 43 (FSR 848); *right panel* open clusters with error bars, respectively.

**Table 5:** The frequency distribution of 3173 stars; Koposov 12 (FSR 802) and 4162 stars; Koposov 43 (FSR 848) open clusters under investigations.

| Koposov 12 (FSR 802) | | | | Koposov 43 (FSR 848) | | | |
|---|---|---|---|---|---|---|---|
| **Intervals (arcmin)** | **Midpoint (arcmin)** | **Frequencies (counts)** | **Density (stars arcmin$^{-3}$)** | **Intervals (arcmin)** | **Midpoint (arcmin)** | **Frequencies (counts)** | **Density (stars arcmin$^{-3}$)** |
| 0.1818 – 0.8228 | 0.5023 | 23 | 9.9857 | 0.1365 – 0.7375 | 0.437 | 31 | 18.6064 |
| 0.8228 – 1.4638 | 1.1433 | 66 | 6.1212 | 0.7375 – 1.3385 | 1.038 | 68 | 8.1466 |
| 1.4638 – 2.1048 | 1.7843 | 93 | 3.5954 | 1.3385 – 1.9395 | 1.639 | 114 | 5.5685 |
| 2.1048 – 2.7458 | 2.4253 | 114 | 2.3972 | 1.9395 – 2.5405 | 2.24 | 157 | 4.1270 |
| 2.7458 – 3.3868 | 3.0663 | 146 | 1.9248 | 2.5405 – 3.1415 | 2.841 | 183 | 2.9972 |
| 3.3868 – 4.0278 | 3.7073 | 156 | 1.4086 | 3.1415 – 3.7425 | 3.442 | 194 | 2.1672 |
| 4.0278 – 4.6688 | 4.3483 | 175 | 1.1494 | 3.7425 – 4.3435 | 4.043 | 215 | 1.7420 |
| 4.6688 – 5.3098 | 4.9893 | 214 | 1.0680 | 4.3435 – 4.9445 | 4.644 | 220 | 1.3516 |
| 5.3098 – 5.9508 | 5.6303 | 221 | 0.8664 | 4.9445 – 5.5455 | 5.245 | 258 | 1.2430 |
| 5.9508 – 6.5918 | 6.2713 | 259 | 0.8186 | 5.5455 – 6.1465 | 5.846 | 259 | 1.0047 |
| 6.5918 – 7.2328 | 6.9123 | 270 | 0.7025 | 6.1465 – 6.7475 | 6.447 | 330 | 1.0527 |
| 7.2328 – 7.8738 | 7.5533 | 288 | 0.6276 | 6.7475 – 7.3485 | 7.048 | 327 | 0.8729 |
| 7.8738 – 8.5148 | 8.1943 | 337 | 0.6241 | 7.3485 – 7.9495 | 7.649 | 353 | 0.8001 |
| 8.5148 – 9.1558 | 8.8353 | 303 | 0.4827 | 7.9495 – 8.5505 | 8.250 | 399 | 0.7775 |
| 9.1558 – 9.7968 | 9.4763 | 385 | 0.5332 | 8.5505 – 9.1515 | 8.851 | 388 | 0.6569 |
| 9.7968 – 10.4378 | 10.1173 | 123 | 0.1494 | 9.1515 – 9.7525 | 9.452 | 459 | 0.6815 |
| – | – | – | – | 9.7525 – 10.3535 | 10.053 | 207 | 0.2717 |
| | | $\Sigma = 3173$ | | | | $\Sigma = 4162$ | |

## 3. Age, reddening, and distance

In this section, we need to determine many of cluster parameters (reddening, distance modulus, ages) due to a photometric analysis by constructing the color-magnitude diagram (CMD) with reduced field star contamination (i.e. established membership). Presently Gaia DR2 catalog will be arranged to download worksheet row data with parallaxes greater than or equal to zero. Therefore, we have 558 stars (4.5 arcmin radius) of Koposov 12 (FSR 802) and 612 stars (4 arcmin radius) of Koposov 43 (FSR 848). Following the consideration devoted by Roeser *et al.* (2010); into which; i) stars with proper motions uncertainties $\geq$ 4.0 (mas/yr) were rejected, ii) moreover to stars with observational uncertainties $\geq$ 0.2 mag (Claria and Lapasset 1986). Therefore the obtained results are 505 stars; Koposov 12 (FSR 802) and 534 stars; Koposov 43 (FSR 848).

Since proper motions play a vital role to separate field stars from the main sequence and to derive authentic fundamental parameters as well (Yadav *et al.* 2013; Sariya *et al.* 2018; Bisht *et al.* 2019; Bisht *et al.* 2020). Figure 4 shows the proper motion vector point diagram (VPD) on both sides with a distribution Histogram of 1.00 (mas/yr) bins. The Gaussian function fit to the central bins provides the mean proper motion in both directions. Therefore iii) all data within the range of mean ($\pm 1\sigma$) (where $\sigma$ is the standard deviation of the mean) can be considered as highly probable astrometric members, i.e. 452 stars; Koposov 12 (FSR 802) and 438 stars; Koposov 43 (FSR 848).

Finally, we used these lists of members and interface those with the PPMXL catalog (Roeser *et al.* 2010) via crossmatch to get a corresponding near a region of the clusters in J, H, and $K_s$ passbands utilizing TOPCAT[3] (Taylor *et al.* 2005) based on The Starlink Tables Infrastructure Library (STIL). Its powerful in our analysis for working with tabular data evaluates a random number in the range $0 \leq x < 1$, and offers many facilities for manipulation of data such as astronomical catalogs. As a result of these above procedures, we have high membership probability $\geq$ 50%, i.e. 285 stars; Koposov 12 (FSR 802), and 310 stars; Koposov 43 (FSR 848).

The cluster parameters (reddening, distance modulus, ages) were estimated with several isochrones of different ages with the theoretical Padova isochrones[4] (Marigo *et al.* 2017) for (J, H, $K_s$, G, $G_{BP}$, $G_{RP}$) colors as in a case of Evans *et al.* (2018). The best fit should be obtained at the same distance modulus for both diagrams, and the reddening of the photometric system by Fiorucci and Munari (2003).

Our fitted color-magnitude diagrams (CMDs) for (J-K, $K_s$), (J-H, J), and ($G_{BP}$-$G_{RP}$, G) mag appear in Figure 5. All numerically astrophysical parameters appear in Table 6 with comments.

Our estimation may be done with Solar metallicity (Z = 0.019) (Froebrich *et al.* 2008) and are found in a region not heavily contaminated by field stars (Bonatto *et al.*

---

[3] http://www.star.bris.ac.uk/~mbt/topcat/
[4] http://stev.oapd.inaf.it/cgi-bin/cmd

2004), we adopt log (age/yr) = 9.00 ± 0.20 for Koposov 12 (FSR 802), which is greater than those obtained with Sampedro *et al.* (2017) and Kharchenko *et al.* (2013) by about log (age/yr) = 0.22 and log (age/yr) = 0.81, respectively; and log (age/yr) = 9.50 ± 0.20 for Koposov 43 (FSR 848), which is also greater than those obtained with same above authors by about log (age/yr) = 0.20 and log (age/yr) = 0.385, respectively.

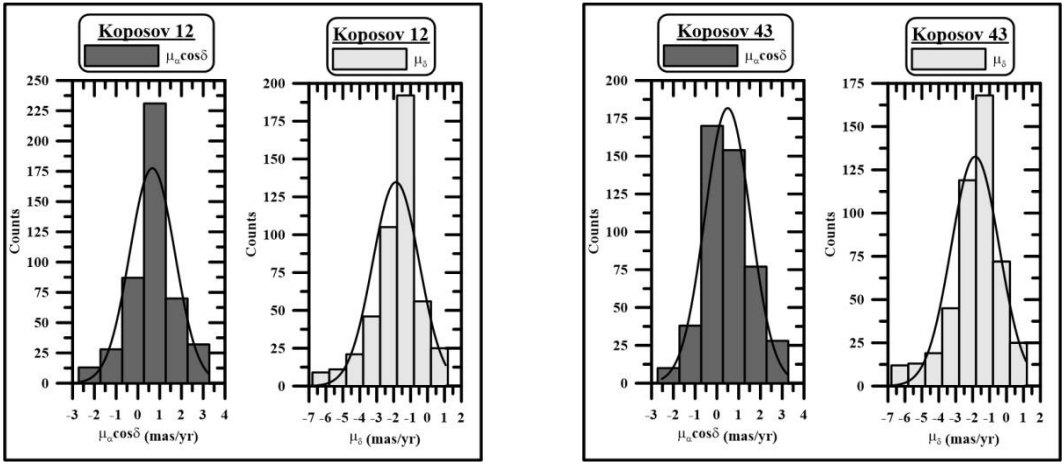

**Fig. 4:** Proper motion vector point diagram (VPD) of proper motions in both directions with a Histogram of 1.00 (mas/yr) bins of Koposov 12 (FSR 802); *left panel* and Koposov 43 (FSR 848); *right panel*. The Gaussian fit to the central bins provides the mean in both directions. The data within the mean range ±1σ can be considered probable astrometric members (candidates).

The reddening (color excess) of the clusters have been determined using the relations of Schlegel *et al.* (1998) and Schlafly and Finkbeiner (2011), where the coefficient ratios $A_J/A_V = 0.276$ and $A_H/A_V = 0.176$ are inferred using absorption ratios by Schlegel *et al.* (1998), whereas the ratio $A_K/A_V = 0.118$ was derived by Dutra *et al.* (2002). For our calculations with (J, H, $K_s$) we used the following results for the color excess of photometric system by Fiorucci and Munari (2003); $E(J-H)/E(B-V) = 0.309 ± 0.130$, $E(J-K_s)/E(B-V) = 0.485 ± 0.150$, where $R_V = A_V/E(B-V) = 3.1$. By using these formulae for these two clusters under examination to rectify the effects of reddening in the color-magnitude diagrams (CMDs) with extinction coefficients $A_V$, i.e. $A_V = 1.407$ and $1.490$, therefore the extinction coefficients for both clusters are $A = 0.276$ and $0.283$ respectively. The line of sight extinction $A_G$ and the reddening $E(G_{BP}-G_{RP})$ are estimated like in Hendy (2018) as $A_G/A_V = 0.859$ and $E(B-V) = 1.289\ E(G_{BP}-G_{RP})$. In this manner we have; $A_G = 1.209$ and $E(G_{BP}-G_{RP}) = 0.352$ for Koposov 12 (FSR 802) and $A_G = 1.280$ and $E(G_{BP}-G_{RP}) = 0.374$ for Koposov 43 (FSR 848).

The most reason for our color-magnitude diagrams (CMDs) demonstrate that the heliocentric distances $[d = 10^{[(m-M)-A+5]/5}; pc]$ is of order about 1850 ± 43 pc and 2500 ± 50 pc, and the reddening E(B-V) are almost 0.454 ± 0.05 and 0.482 ± 0.05 mag for Koposov 12 (FSR 802) and Koposov 43 (FSR 848), respectively.

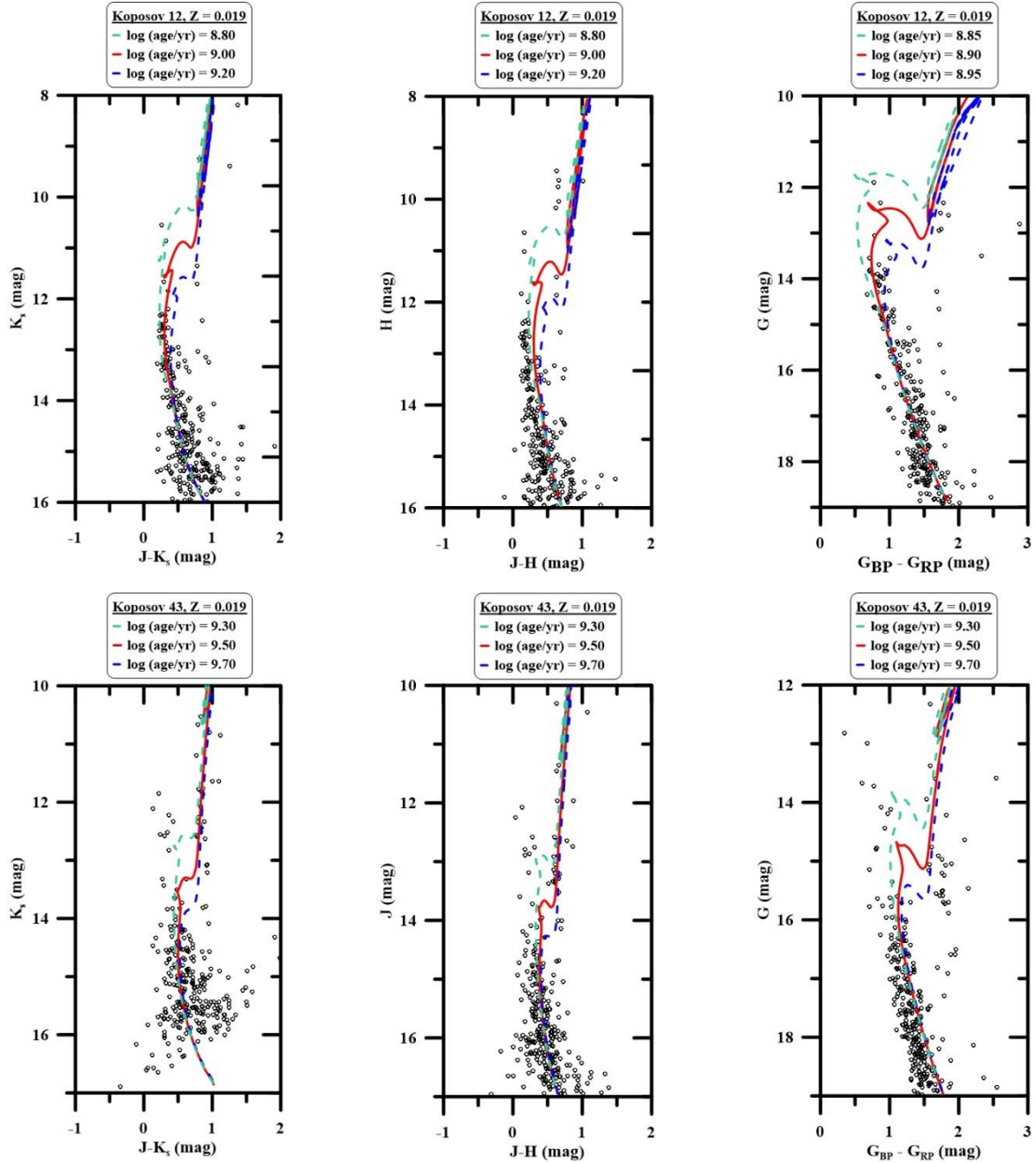

**Fig. 5:** Padova color-magnitude diagram (CMD) isochrones over [$K_s$ *vs.* (J-$K_s$), J *vs.* (J-H), and G *vs.* ($G_{BP}$-$G_{RP}$)] with Evans *et al.* (2018) for Koposov 12 (FSR 802); *upper panel* and Koposov 43 (FSR 848); *lower panel*.

Table 6: Comparison between our photometric parameters of Koposov 12 (FSR) and Koposov 43 (FSR 848) open clusters with different authors.

| Parameters | Koposov 12 (FSR 802) | Koposov 43 (FSR 848) | References |
|---|---|---|---|
| log (age/yr) | 9.00 ± 0.20 | 9.50 ± 0.20 | Present study |
|  | 8.78 | 9.30 | Sampedro *et al.* (2017) |
|  | 8.190 | 9.115 | Kharchenko *et al.* (2013) |
|  | 8.8 | - | Yadav *et al.* (2011) |
|  | 9.00 | 9.30 | Froebrich *et al.* (2008) |
| d (pc) | 1850 ± 43 | 2500 ± 50 | Present study |
|  | 2351.2 | 4787.5 | Soubiran *et al.* (2018) |
|  | 2525.25 | 5555.56 | Cantat-Gaudin *et al.* (2018) |
|  | 2000 | 2800 | Sampedro *et al.* (2017) |
|  | 1900 | 3000 | Kharchenko *et al.* (2013) |
|  | 2000 ± 200 | - | Yadav *et al.* (2011) |
|  | 2050 | 2800 | Froebrich *et al.* (2008) |
| E(B-V) | 0.454 ± 0.05 | 0.482 ± 0.05 | Present study |
|  | 0.510 | 0.380 | Sampedro *et al.* (2017) |
|  | 0.450 | 0.650 | Kharchenko *et al.* (2013) |
|  | 0.51 ± 0.05 | - | Yadav *et al.* (2011) |
| E(J-$K_s$) | 0.220 ± 0.06 | 0.233 ± 0.06 | Present study |
|  | 0.216 | 0.312 | Kharchenko *et al.* (2013) |
| E(J-H) | 0.140 ± 0.08 | 0.149 ± 0.07 | Present study |
|  | 0.144 | 0.208 | Kharchenko *et al.* (2013) |
| (m-M) | 11.60 ± 0.28 | 12.25 ± 0.28 | Present study |
|  | 11.55 ± 0.03 | 12.21 ± 0.09 | Koposov *et al.* (2008) |

## 4. Luminosity and mass functions

In this section, we estimate the luminosity function (LF) and mass function (MF) of the clusters. We have got good results in the photometric parameters and position of the clusters. Hence, we will induce their luminosity function (LF); describes the total number of stars in different absolute magnitudes, and mass function (MF). Figure 6 presents luminosity functions (LFs) of Koposov 12 (FSR 802) and Koposov 43 (FSR 848) with absolute $K_s$ magnitude in the range -3.415 < $M_{Ks}$ < 5.658 and -3.234 < $M_{Ks}$ < 4.982 respectively. The total luminosities have computed for both clusters with values of 2.84 ± 1.37 mag and 2.57 ± 1.33 mag respectively. In mass-segregation massive stars are concentrated towards the cluster core than fainter ones and this phenomenon has been reported recently for many open clusters (Piatti 2016, Zeidler *et al.* 2017; Dib *et al.* 2018, Rangwal *et al.* 2019, Bisht *et al.* 2020, Joshi *et al.* 2020).

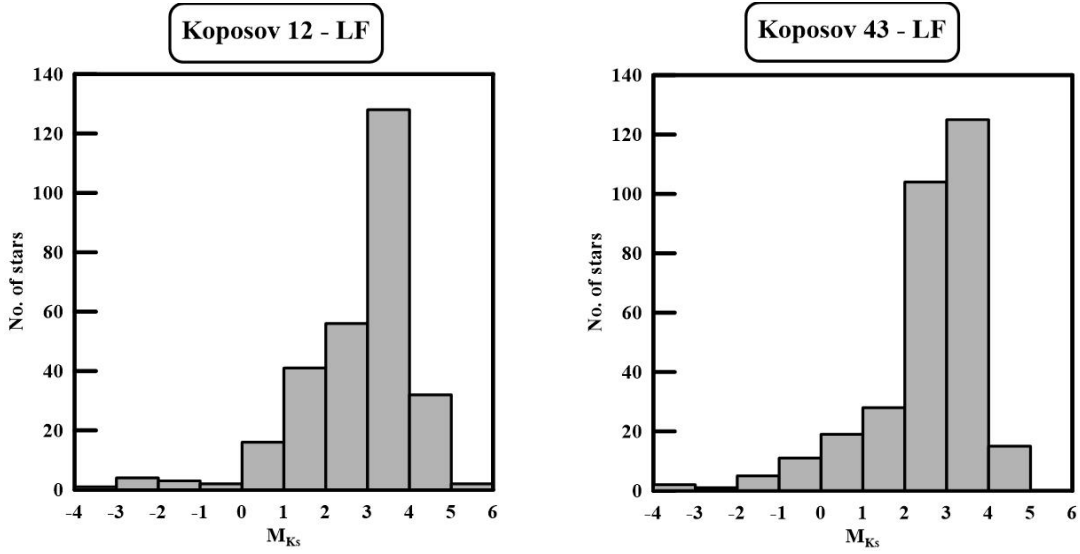

**Fig. 6:** *Left panel* presents the luminosity function (LF) of Koposov 12 (FSR 802), while the *right panel* shows that for Koposov 43 (FSR 848) open clusters.

In arrange and to compute the mass function (MF) of these open clusters; we have characterized the initial mass function (IMF) with power-law as takes like:

$$\frac{dN}{dM} \propto M^{-\Gamma}. \tag{5}$$

where (dN/dM) is the number of stars on the mass interval $(M: M + dM)$ and $\Gamma$ is a dimensionless exponent. From Salpeter (1955), the initial mass function (IMF) for massive stars ($> 1 M_\odot$) has been considered and well built-up (i.e. $\Gamma = 2.35$).

According to well-known mass-luminosity relation (MLR), and accounting to absolute magnitudes $M_{K_s}$ and masses $(M/M_\odot)$ of the adopted isochrones (Evans *et al.* 2018), we can infer the mass-luminosity relation (MLR) of each cluster (Elsanhoury and Nouh 2019). This relationship is polynomial functions of the $2^{nd.}$ order for two ranges of luminosities, as shown in Figure 7. Therefore;

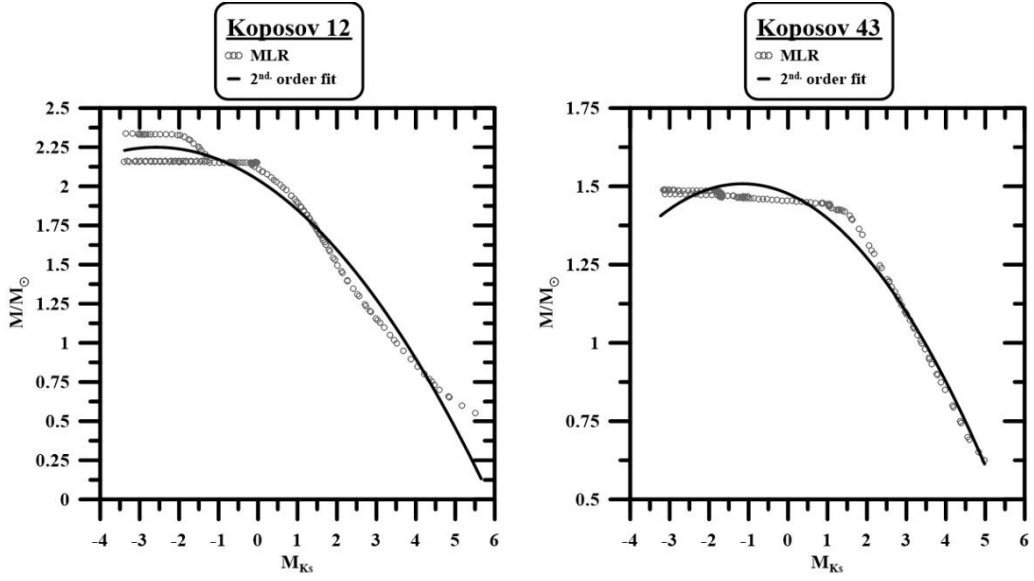

**Fig. 7:** Mass-luminosity relation (MLR) between absolute magnitude $M_{K_s}$ and masses ($M/M_\odot$) with Solar metallicity (Z = 0.019) *(gray circles)* from isochrones (Evans *et al.* 2018) and it's fitted lines *(black solid)* for Koposov 12 (FSR 802); *left panel* and Koposov 43 (FSR 848); *right panel*.

- For Koposov 12 (FSR 802);

$$[M/M_\odot]_{Koposov\ 12} = 2.042 - 0.162\ M_{K_s} - 0.032 M_{K_s}^2, \qquad (6)$$

- For Koposov 43 (FSR 848);

$$[M/M_\odot]_{Koposov\ 43} = 1.476 - 0.055\ M_{K_s} - 0.024 M_{K_s}^2. \qquad (7)$$

Then, the total masses are; 364 ± 19 $M_\odot$ and 352 ± 19 $M_\odot$ for Koposov 12 and Koposov 43, respectively.

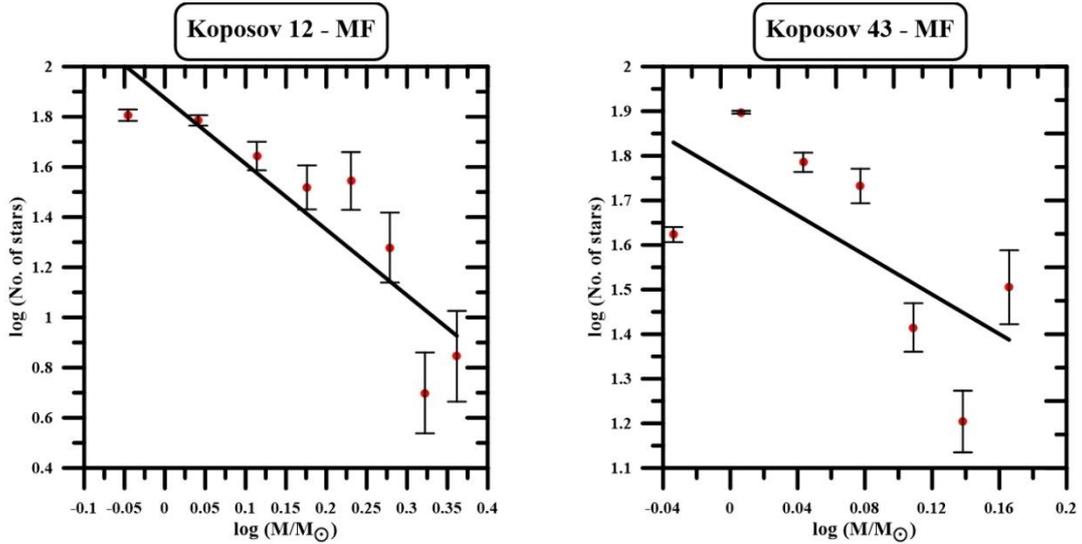

**Fig. 8:** The mass functions (MFs) of both Koposov 12 (FSR 802); *left panel* and Koposov 43 (FSR 848); *right panel* with their fitted lines.

Figure 8 presents the mass functions (MFs) of these open clusters with error bars, appearing its fitted line with slopes -2.62 ± 0.56; Koposov 12 and -2.22 ± 0.90; Koposov 43, which are in good agreement with Salpeter's value (1955).

**5. Dynamical and kinematical structure**

We are going presently to consider the dynamical and kinematical processes take place of these two open clusters, while the obtained results have shown up in Table 7 with comments.

It's known that in the Galactic disk, the effect on the gravitational massive body (e.g. star cluster) is given by (Röser and Schilbach 2019) formula, i.e.

$$x_L = \left[\frac{GM_C}{4A(A-B)}\right]^{1/3} = \left[\frac{GM_C}{4\Omega_o^2 - \kappa^2}\right]^{1/3} \qquad (8)$$

where ($x_L$) is the gap of the Lagrangian points within the center, ($M_C$) is the sum of the collective mass (with Equations 6 and 7), toward the gap from the center, which mentioned to the tidal radius $r_t$ of the cluster (i.e. $x_L \approx r_t$); into which the star undergoes equal forces due to gravitational pull toward the cluster and in opposite direction (i.e. the Galactic center), ($\Omega_o = A - B$) the angular velocity, and ($\kappa = \sqrt{-4B(A-B)}$) the epicyclic frequency at the position of the Sun (both in km s-1 kpc-1) (Röser et al. 2011), (A) and (B) are Oort's constants; 15.3 ± 0.4 km s$^{-1}$ kpc$^{-1}$ and -11.9 ± 0.4 km s$^{-1}$ kpc$^{-1}$ with Bovy (2017) and also equals to 15.6 ± 1.6 km s$^{-1}$ kpc$^{-1}$ and -13.9 ± 1.8 km s$^{-1}$ kpc$^{-1}$ by (Nouh & Elsanhoury 2020), and G = 4.30 × 10$^{-6}$ kpc/M$_\odot$ (km s$^{-1}$)$^2$ is the Gravitational constant. In such a way; our obtained $r_t \approx 9.80 \pm 3.13$ pc as a function of our total estimated masses ($M_C$) for Koposov 12 (FSR 802) and 9.70 ± 3.11 pc for Koposov 43 (FSR) clusters.

Due to forces of contraction and/or destruction, open clusters come to Maxwellian stability equilibrium with a time characterized as a relaxation time ($T_{relax}$), into which the cluster will lose all traces of its initial dynamic condition (Yadav *et al.* 2013; Bisht *et al.* 2019). $T_{relax}$ depends on dynamical crossing time ($T_{cross}$) and the number (N) of member stars (Lada and Lada, 2003). During relaxation time ($T_{relax}$) low mass stars possess the largest random velocity, possessing a bigger volume than the high mass does (Mathieu and Latham, 1986). Mathematically relaxation time ($T_{relax}$) has the form.

$$T_{relax} = \frac{N}{8 \ln N} T_{cross}. \qquad (9)$$

where $(T_{cross} = D/\sigma_V)$ is the dynamical crossing time; namely defined as the time needed (independent of the size and shape of the orbit) for the cluster to be cross the Galaxy once (Binney and Tremaine, 1998). Typically crossing time $(T_{cross})$ in open clusters $\sim 10^6$ years (Lada and Lada, 2003), (D) is the cluster diameter (Maciejewski and Niedzielski 2007) with expression corresponds to Lada and Lada (2003) as $(D \simeq 2r_{lim})$. All numerical values of crossing time $(T_{cross})$ and relaxation time $(T_{relax})$ are drawn here with Table 7.

For enough cluster members, the time needed to eject all its members from internal stellar encounters defined as the evaporation time $(\tau_{ev})$ (Adams and Myers, 2001) to be greater than $10^8$ years; $(\tau_{ev})$ for a stellar system in virial equilibrium is of order $10^2$ $(T_{relax})$. For the cluster to remain bound in the face, the escaping velocity $(V_{esc})$ of rapid gas removal from the cluster is $\left(V_{esc} = R_{gc}\sqrt{2GM_C/r_{lim}}\right)$ and must be less than the dispersion velocity $(\sigma_V)$ (Lada and Lada, 2003). Thus a bound group will emerge only if the star-formation efficiency (SFE, characterizes most cluster-forming dense cores) is greater than 50% (Wilking and Lada, 1983).

Finally, we can describe the dynamical state of these clusters by computing their dynamical evolution parameter $(\tau = age/T_{relax})$. If the cluster age founded greater than its relaxation time; i.e. $\tau \gg 1$ then the cluster was dynamically relaxed and vice versa.

Now, we focused on some kinematics and velocity ellipsoid parameters (VEPs) for our objects with member (N) stars due to the computational algorithm presented by (Elsanhoury *et al.* 2015, 2018; Elsanhoury 2015; Elsanhoury 2016; Postnikova *et al.* 2020; Bisht *et al.* 2020), with considering members coordinated with $(\alpha, \delta)$ located at a distance $d_i$ (pc), proper motions (mas/yr) in both directions, and radial velocities $V_r$ (km/s) which are listed in Table 1. Therefore, and according to well-known basic equations governs the positions $(X, Y, Z; pc)$ from the Sun along with the equatorial system and distance d (pc) of the star members (Mihalas and Binney 1981), i.e.

$$X = d \cos \delta \cos \alpha, \qquad (10)$$
$$Y = d \cos \delta \sin \alpha, \qquad (11)$$
$$Z = d \sin \delta. \qquad (12)$$

In this manner, differentiating Equations (10, 11, and 12) with respect to time, we get the velocity components $(V_X, V_Y, V_Z; \text{km s}^{-1})$ along $X, Y, \text{and } Z - \text{axes}$ in the coordinate system with respect to the Sun (Smart 1968).

i.e.

$$V_X = -4.74 d_i \mu_\alpha \cos \delta \sin \alpha - 4.74 d_i \mu_\delta \sin \delta \cos \alpha + V_r \cos \delta \cos \alpha, \tag{13}$$

$$V_Y = +4.74 d_i \mu_\alpha \cos \delta \cos \alpha - 4.74 d_i \mu_\delta \sin \delta \sin \alpha + V_r \cos \delta \sin \alpha, \tag{14}$$

$$V_Z = +4.74 d_i \mu_\delta \cos \delta + V_r \sin \delta. \tag{15}$$

To get the components of space velocities $(U, V, W)$ along with Galactic coordinates as a function of space stellar velocities $(V_X, V_Y, V_Z)$ whose definite to the Sun were derived Liu *et al.* (2011); i.e. from the equatorial to the Galactic coordinates, based on NIR by 2MASS (Skrutskie *et al.* 2006) and radio observation data.

i.e.

$$U = -0.0518807421\, V_X - 0.872222642\, V_Y - 0.4863497200\, V_Z, \tag{16}$$

$$V = +0.4846922369\, V_X - 0.4477920852\, V_Y + 0.5713692061 V_Z, \tag{17}$$

$$W = -0.8731447899\, V_X - 0.1967483417\, V_Y + 0.4459913295\, V_Z, \tag{18}$$

A brief description of the algorithm mentioned above will deliver here. Let $(\xi)$ and it's zero points focuses coincides with the center of the distribution and let $(l, m, \text{and } n)$ be the direction cosines of the axis with respect to the shifted one, then the coordinates $(Q_i)$ of the point i, with respect to the $(\xi)$ axis, are given by

$$Q_i = l(U_i - \overline{U}) + m(V_i - \overline{V}) + n(W_i - \overline{W}). \tag{19}$$

where $(\overline{U}, \overline{V}, \overline{W})$ are the mean velocities, and considering $(\sigma^2)$ may be a generalization of the mean square deviation, i.e.

$$\sigma^2 = \frac{1}{N} \sum_{i=1}^{N} Q_i^2 \tag{20}$$

Due to mean velocities $(\overline{U}, \overline{V}, \overline{W})$, and Equations (19) and (20) we deduce after some calculations that

$$\sigma^2 = \underline{x}^T B \underline{x} \tag{21}$$

Where $(\underline{x})$ is the $(3 \times 3)$ direction cosines vector, and $(B)$ is $(3 \times 3)$ symmetric matrix elements $(\mu_{ij})$

and

$$\mu_{11} = \frac{1}{N}\sum_{i=1}^{N} U_i^2 - (\overline{U})^2; \quad \mu_{12} = \frac{1}{N}\sum_{i=1}^{N} U_i V_i - \overline{U}\overline{V};$$

$$\mu_{13} = \frac{1}{N}\sum_{i=1}^{N} U_i W_i - \overline{U}\overline{W}; \quad \mu_{22} = \frac{1}{N}\sum_{i=1}^{N} V_i^2 - (\overline{V})^2; \tag{22}$$

$$\mu_{31} = \frac{1}{N}\sum_{i=1}^{N} V_i W_i - \overline{V}\overline{W}; \quad \mu_{33} = \frac{1}{N}\sum_{i=1}^{N} W_i^2 - (\overline{W})^2;$$

Now, the necessary conditions for an extremum are

$$(B - \lambda I)\underline{x} = 0 \tag{23}$$

These are three homogenous equations in three unknowns have a nontrivial solution if and only if

$$D(\lambda) = |B - \lambda I| = 0, \tag{24}$$

The above equation is the characteristic equation for the matrix (B), where ($\lambda$) is the eigenvalue, ($\underline{x}$) and (B) could be written as

$$\underline{x} = \begin{bmatrix} l \\ m \\ n \end{bmatrix} \text{ and } B = \begin{vmatrix} \mu_{11} & \mu_{12} & \mu_{13} \\ \mu_{12} & \mu_{22} & \mu_{23} \\ \mu_{13} & \mu_{23} & \mu_{33} \end{vmatrix}$$

Then the required roots (i.e. eigenvalues) are

$$\lambda_1 = 2\rho^{1/3}\cos\frac{\phi}{3} - \frac{k_1}{3};$$

$$\lambda_2 = -\rho^{1/3}\left\{\cos\frac{\phi}{3} + \sqrt{3}\sin\frac{\phi}{3}\right\} - \frac{k_1}{3}; \tag{25}$$

$$\lambda_3 = -\rho^{1/3}\left\{\cos\frac{\phi}{3} - \sqrt{3}\sin\frac{\phi}{3}\right\} - \frac{k_1}{3}$$

where

$$k_1 = -(\mu_{11} + \mu_{22} + \mu_{33}),$$

$$k_2 = \mu_{11}\mu_{22} + \mu_{11}\mu_{33} + \mu_{22}\mu_{33} - (\mu_{12}^2 + \mu_{13}^2 + \mu_{23}^2), \tag{26}$$

$$k_3 = \mu_{12}^2\mu_{33} + \mu_{13}^2\mu_{22} + \mu_{23}^2\mu_{11} - \mu_{11}\mu_{22}\mu_{33} - 2\mu_{12}\mu_{13}\mu_{23}.$$

$$q = \frac{1}{3}k_2 - \frac{1}{9}k_1^2; \quad r = \frac{1}{6}(k_1 k_2 - 3k_3) - \frac{1}{27}k_1^3 \tag{27}$$

$$\rho = \sqrt{-q^3} \tag{28}$$

$$x = \rho^2 - r^2 \tag{29}$$

and

$$\phi = \tan^{-1}\left(\frac{\sqrt{x}}{r}\right) \tag{30}$$

Depending on the matrix that controls the eigenvalue problem [Equation (24)] for the velocity ellipsoid, we built up analytical expressions of some parameters in terms of the $(3 \times 3)$ matrix elements $(\mu_{ij})$. Table 7 shows all those numerical results.

- ***The direction cosines parameters***

The direction cosines $(l_j, m_j, n_j; \forall j = 1,2,3)$ for the eigenvalue problem $(\lambda_j)$, matrix elements $(\mu_{ij})$, and dispersion velocities $(\sigma_j)$ [i.e. $\sigma_j = \sqrt{\lambda_j}; \forall j = 1,2,3$] along three axes (Elsanhoury *et al.* 2015) are mathematically given by the following:

$$l_j = [\mu_{22}\mu_{33} - \sigma_j^2(\mu_{22} + \mu_{33} - \sigma_j^2) - \mu_{23}^2]/D_j, \tag{31}$$

$$m_j = [\mu_{23}\mu_{13} - \mu_{12}\mu_{33} + \sigma_j^2\mu_{12}]/D_j, \tag{32}$$

$$n_j = [\mu_{12}\mu_{23} - \mu_{13}\mu_{22} + \sigma_j^2\mu_{13}]/D_j, \tag{33}$$

where $(l_j^2 + m_j^2 + n_j^2 = 1)$ as an initial test for our code for our sample and,

$$D_j^2 = (\mu_{22}\mu_{33} - \mu_{23}^2)^2 + (\mu_{23}\mu_{13} - \mu_{12}\mu_{33})^2 + (\mu_{12}\mu_{23} - \mu_{13}\mu_{22})^2 + 2[(\mu_{22} + \mu_{33})(\mu_{23}^2 + \mu_{22}\mu_{33}) + \mu_{12}(\mu_{23}\mu_{13} - \mu_{12}\mu_{33}) + \mu_{13}(\mu_{12}\mu_{23} - \mu_{13}\mu_{22})]\sigma_j^2 + (\mu_{33}^2 + 4\mu_{22}\mu_{33} + \mu_{22}^2 - 2\mu_{23}^2 + \mu_{12}^2 + \mu_{13}^2)\sigma_j^4 - 2(\mu_{22} + \mu_{33})\sigma_j^6 + \sigma_j^8.$$

- ***The Galactic longitude and Galactic latitude parameters***

Let $(L_j)$ and $(B_j)$, $(\forall j = 1,2,3)$ be the Galactic longitude and the Galactic latitude of the directions, respectively which correspond to the extreme values of the dispersion, then

$$L_j = \tan^{-1}\left(\frac{-m_j}{l_j}\right), \tag{34}$$

$$B_j = \sin^{-1}(n_j). \tag{35}$$

- *The center of the cluster*

    The center of the cluster $(x_c, y_c, z_c)$ can be derived by the simple method of finding the equatorial coordinates of the center of mass for the number $(N_i)$ of discrete objects, i.e.

$$x_c = \left[\sum_{i=1}^{N} d_i \cos\alpha_i \cos\delta_i\right] / N, \tag{36}$$

$$y_c = \left[\sum_{i=1}^{N} d_i \sin\alpha_i \cos\delta_i\right] / N, \tag{37}$$

$$z_c = \left[\sum_{i=1}^{N} d_i \sin\delta_i\right] / N. \tag{38}$$

- *Projected distances*

    Considering our estimated distances d (pc) with over markers, consequently, we infer to include those distances to the Galactic center ($R_{gc}$) (Mihalas and Binney 1981) like a function of the Sun's distance from the Galactic center (i.e. $R_o = 8.20 \pm 0.10$ kpc) as mentioned recently with Bland-Hawthorn *et al.* (2019); i.e. $R_{gc}^2 = R_o^2 + d^2 - 2R_o d \cos l$, in such a way the anticipated (projected) distances towards the Galactic plane $(X_\odot, Y_\odot)$ and the distance from the Galactic plane $(Z_\odot)$ (Tadross 2011) may be computed as:

$$X_\odot = d\cos b \cos l, \ Y_\odot = d\cos b \sin l, \text{ and } Z_\odot = d\sin b \tag{39}$$

- *The Solar elements*

    Consider a group with spatial velocities $(\overline{U}, \overline{V}, \overline{W})$. The components of the Sun's velocities are $(U_\odot, V_\odot, \text{and } W_\odot)$ are given as; $(U_\odot = -\overline{U})$, $(V_\odot = -\overline{V})$, and $(W_\odot = -\overline{W})$. Therefore, we have the Solar elements with spatial velocities considered w.s.v.c. like;

$$S_\odot = \sqrt{\overline{U}^2 + \overline{V}^2 + \overline{W}^2}, \tag{40}$$

$$l_A = \tan^{-1}\left(\frac{-\overline{V}}{\overline{U}}\right), \tag{41}$$

$$b_A = \sin^{-1}\left(\frac{-\overline{W}}{S_\odot}\right). \tag{42}$$

Now consider the position along X, Y, and Z − axes in the coordinate system whose centered at the Sun, then the Sun's velocities with respect to this same group and referred to the same axes are given as; $(X_\odot^\bullet = -\overline{V}_X), (Y_\odot^\bullet = -\overline{V}_Y),$ and $(Z_\odot^\bullet = -\overline{V}_Z)$. Therefore, we have obtained the Solar elements with radial velocities considered w.r.v.c. as;

$$S_\odot = \sqrt{(X_\odot^\bullet)^2 + (Y_\odot^\bullet)^2 + (Z_\odot^\bullet)^2}, \tag{43}$$

$$\alpha_A = \tan^{-1}\left(\frac{Y_\odot^\bullet}{X_\odot^\bullet}\right), \tag{44}$$

$$\delta_A = \tan^{-1}\left(\frac{Z_\odot^\bullet}{\sqrt{(X_\odot^\bullet)^2 + (Y_\odot^\bullet)^2}}\right). \tag{45}$$

where $(l_A, \alpha_A)$ is the Galactic; longitude and right ascension of the Solar apex and $(b_A, \delta_A)$ are the Galactic; latitude and declination of the Solar apex, and $(S_\odot)$ is considered as the absolute value of the Sun's velocity relative to our groups under investigations.

**Table 7:** Our dynamical and kinematical parameters of Koposov 12 (FSR 802) and Koposov 43 (FSR 848) open clusters.

| Parameters | Koposov 12 (FSR 802) | Koposov 43 (FSR 848) | References |
|---|---|---|---|
| No. of members (N) | 285 | 310 | Present study |
| $\mu_\alpha \cos\delta$ (mas/yr) | 0.632 ± 0.006 | 0.517 ± 0.057 | Present study |
| $\mu_\delta$ (mas/yr) | -1.945 ± 0.006 | -1.810 ± 0.057 | Present study |
| Luminosity (mag) | 2.84 ± 1.37 | 2.57 ± 1.33 | Present study |
| $\Gamma$ | 2.62 ± 0.56 | 2.22 ± 0.90 | Present study |
| Total mass $M_c$ ($M_\odot$) | 364 ± 19 | 352 ± 19 | Present study |
| Average mass ($M_\odot$) | 1.276 | 1.138 | Present study |
| $r_t$ (pc) | 9.80 ± 3.13 | 9.70 ± 3.11 | Present study |
| $T_{cross}$ (Myr) | 10.316 ± 3.22 | 12.750 ± 3.57 | Present study |
| $T_{relax}$ (Myr) | 65.017 ± 8.06 | 86.125 ± 9.28 | Present study |
| $\tau_{ev}$ (Myr) | 6501 ± 80.67 | 8613 ± 92.81 | Present study |
| $\tau$ | 15.38 ± 3.92 | 36.72 ± 6.06 | Present study |
| $V_{esc}$ (km/s) | 311 ± 5.67 | 332 ± 5.48 | Present study |
| $(\overline{V}_X, \overline{V}_Y, \overline{V}_Z)$ (km/s) | -15.62 ± 3.95, 32.50 ± 5.70, -22.50 ± 4.74 | -31.17 ± 5.58, 55.68 ± 7.46, -41.56 ± 6.45 | Present study |
| $(\overline{U}, \overline{V}, \overline{W})$ (km/s) | -16.60 ± 4.07, -39.03 ± 6.25, -2.80 ± 0.60 | -26.74 ± 5.17, -71.27 ± 8.44, -2.27 ± 0.66 | Present study |
| | -21.81 ± 1.44, -19.24 ± 0.29, -0.46 ± 0.19 | 2.22 ± 1.88, -32.09 ± 1.73, -20.01 ± 1.14 | Soubiran et al. (2018) |
| $(\lambda_1, \lambda_2, \lambda_3)$ (km/s) | 13841.1, 827.712, 134.462 | 236662, 8411.11, 608.45 | Present study |
| $(\sigma_1, \sigma_2, \sigma_3)$ (km/s) | 117.648, 28.77, 11.596 | 486.48, 91.71, 24.67 | Present study |
| $\sigma_V$ (km/s) | 122 ± 9.00 | 496 ± 4.50 | Present study |
| $(l_1, m_1, n_1)^o$ | 0.088, 0.983, -0.158 | 0.002, -0.921, 0.389 | Present study |
| $(l_2, m_2, n_2)^o$ | -0.313, -0.123, -0.942 | -0.129, -0.386, -0.913 | Present study |
| $(l_3, m_3, n_3)^o$ | 0.946, -0.133, -0.297 | 0.992, -0.048, -0.120 | Present study |
| $(x_c, y_c, z_c)$ (pc) | -17.074, 3792.7, 2683.33 | 334.563, 10042.5, 5787.05 | Present study |
| $B_j, j = 1, 2, 3$ | -9°.100, -70°.359, -17°.253 | 22°.902, -65°.977, -6°.865 | Present study |
| $L_j, j = 1, 2, 3$ | -84°.865, 158°.471, -172°.014 | 89°.589, 108°.446, -177°.226 | Present study |
| $X_\odot$ (kpc) | -1.836 ± 0.043 | -2.499 ± 0.050 | Present study |
| | -2.333 | -4.7852 | Cantat-Gaudin et al. (2018) |
| $Y_\odot$ (kpc) | 0.124 ± 0.001 | 0.004 ± 0.0002 | Present study |
| | 0.1568 ± 0.002 | 0.006 ± 0.0003 | Soubiran et al. (2018) |
| | 0.1568 | 0.0064 | Cantat-Gaudin et al. (2018) |
| $Z_\odot$ (kpc) | 0.194 ± 0.0014 | 0.076 ± 0.009 | Present study |
| | 0.2603 ± 0.0029 | 0.1599 ± 0.0077 | Soubiran et al. (2018) |
| | 0.2463 | 0.1459 | Cantat-Gaudin et al. (2018) |
| $R_{gc}$ (kpc) | 9.347 ± 0.097 | 10.00 ± 0.100 | Present study |
| | 10.6742 | 13.1252 | Cantat-Gaudin et al. (2018) |
| | 10.50 | - | Yadav et al. (2011) |
| | 10.00 | 10.90 | Froebrich et al. (2008) |
| $S_\odot$ (km/s) | 42.50 | 76.15 | Present study |
| $(l_A, b_A)^o$ w.s.v.c. | 66.96, 25.67 | 69.44, 29.24 | Present study |
| $(\alpha_A, \delta_A)^o$ w.r.v.c | 271.10, 40.27 | 267.10, 43.22 | Present study |

## 6. Conclusion

In the present paper we have investigated the poorly studied open clusters; Koposov 12 (FSR 802) and Koposov 43 (FSR 848) using PPMXL and Gaia DR2 data via crossmatch. Here, we re-estimated the cluster center and core radius based on their radial density profile.

To derive their fundamental and kinematical parameters within (J, H, $K_s$, G, $G_{BP}$, $G_{RP}$) regions, we have estimated their membership (utilizing proper motions and the magnitude uncertainties) with probabilities ($\geq 50\%$), and have found 285 and 310 member stars respectively.

Based on the NIR and Gaia CMDs of the cluster members, we have estimated the cluster parameters (reddening, distance modulus, ages) listed in Table 6, which are in good agreement with previous studies. The dynamical and kinematical properties of these clusters (tidal radii, crossing, relaxation and evaporation times, space velocities, and velocity ellipsoid parameters (VEPs)) are numerically mentioned in Table 7.

Based on estimated dynamical evolution parameter ($\tau \gg 1$); i.e. $\tau = 15.38 \pm 3.92$ for Koposov 12 (FSR 802) and $36.72 \pm 6.06$ for Koposov 43 (FRS 848), we infer that these clusters are dynamically relaxed.

*Acknowledgments*

The author is thankful to the referees of this paper for many useful comments and continuous encouragement which highly improved the level of this paper.

This work presents results from the European Space Agency (ESA) space mission Gaia. Gaia data are being processed by the Gaia Data Processing and Analysis Consortium DPAC. Funding for the DPAC is provided by national institutions, in particular, the institutions participating in the Gaia Multi-Lateral Agreement MLA. The Gaia mission website is https://www.cosmos.esa.int/gaia The Gaia Archive website is http://archives.esac.esa.int/gaia. Also, it is worthy to mention that, this publication made use of the data products from the PPMXL catalog.

The author gratefully acknowledges the approval and the support of this research study by the grant number SAR-2018-3-9-F-7591 from the Deanship of Scientific Research at Northern Border University, Arar, Saudi Arabia.